\documentstyle[aps,floats,prl,twocolumn]{revtex}

\begin{document}
\title{On decoherence in quantum algorithm via dynamic models for quantum
measurement}
\author{C.P Sun and H.Zhan}
\address{Institute of Theoretical Physics, Academia Sinica,Beijing 100080, China}
\author{X.F.Liu}
\address{Department of Mathematics, Peking University, Beijing 100871, China}
\date{\today}
\maketitle

\begin{abstract}
The possible effect of environment on the efficiency of a quantum algorithm
is considered explicitely. It is illustrated through the example of Shor's
prime factorization algorithm that this effect may be disastrous. The
influence of environment on quantum computation is probed on the basis of
its analogy to the problem of wave function collapse in quantum
measurement.Techniques from the Hepp-Colemen approach and its generalization
are used to deal with decoherence problems in quantum computation including
dynamic mechanism of decoherence , quantum error avoiding tricks and
calculation of decoherence time.
\end{abstract}


\quad

{\bf PACS number(s)}: 03.65-BZ, 71.10.Ad,89.70.+c

\qquad

{\bf 1. Introduction }

\qquad

Quantum computations (QC) can be understood as a quantum- mechanical time
evolution of certain quantum systems (so-called qubits) [1-9], in which the
non-classical dynamic feature, such as the quantum coherence of states,
plays a dominant role. Indeed, it is the purely quantum characters that
makes it possible for a theoretical quantum computer to solve certain hard
mathematical problems efficiently. In this respect, perhaps the most
important example is Shor's prime factorization algorithm [6]. As the
quantum computation is a quantum process, preserving coherence, at least to
some extent, throughout the whole process is thus an essential requirement.
In fact, the decoherence resulting from the coupling with environment may
make an quantum algorithm invalid and may cause unwelcome exponential
increase of errors in output results [10-12]. Actually, the decoherence
process was even regarded as a mechanism for enforcing classical behavior in
the macroscopic realm [13]. In this view a decohered quantum computer
becomes a classical one. To overcome this difficulty caused by decoherence,
some schemes have been proposed in the last several years [14-20]. Among
them are the quantum error correcting technique inspired by the classical
error correction theory and the schemes avoiding decoherence presented in
connection with the strategy preventing decoherence from certain subsets of
quantum states [21-23]. The latter was systematically described in the
framework of error avoiding quantum coding (EAQC). For the mathematical
details we refer the readers to ref.[23].

It is recongnized that there exists a substantially close relation between
the problem of decoherence in quantum computation and the problem of wave
function collapse(WFC, also called von Neumann's reduction) in quantum
measurement [24]. In the view point of quantum dynamics [25-35] for quantum
measurement, both the measured system and the measuring instrument
(detector) obey the Schrodinger equation and the dynamic evolution governed
by their interaction can result in the WFC under certain conditions. For
example, the collapse happens if the detector contains a great number of
particles or if the detector is in a state with a very large quantum
number.These two cases are usually referred to as macroscopic limit and
classical limit respectively [30]. We recall that in the traditional theory
of quantum measurement [24], the WFC postulate is only an extra assumption
added to the ordinary quantum mechanics. Under this postulate, once we
measure an observable and obtain a definite value $a_k$ the state of the
system must collapse into the corresponding eigenstate $|k\rangle $ from a
coherent superposition $|\phi \rangle =\sum_kc_k|k\rangle \langle k|$. In
the terminology of density matrix this process can be described by a
projection $\rho =|\phi \rangle \langle \phi |\rightarrow $ $\hat \rho =\sum
|c_k|^2|k\rangle \langle k|$ from a pure state to a mixed state. This
projection process means the loss of quantum coherence. There is a strong
resemblance between this process and the quantum decoherence of a quantum
computer resulting from the coupling with the surrounding environment.Thus,
with the correspondence of the environment surrounding the quantum computer
to the measurement instrument monitoring the measured system, some obtained
results in the quantum dynamic models [26-36,] based on Hepp-Colemen (HC)
approach [25] for quantum measurement can be applied to discuss decoherence
problems in quantum computation such as the dynamic mechanism of decoherence
in quantum computer, the quantum error avoiding techniques and the
calculation of the decoherence time of a quantum computation process. We can
also re-consider the strategy grouping the quantum states of qubits to form
decoherence free subsets [21-23] and analyse the dynamic process decohering
the qubits beyond the decoherence free subsets .

In comparison with the dynamic theory of quantum measurement, the system of
qubits in quantum computation is an open system $S$ (similar to the measured
system) surrounded by an environment ( similar to the measuring
instrument,detector). The environment may cause two unwelcome effects on
computation process, namely,causing the states of the qubit system to
continuously decohere to approach classical states [13, 25-36]and
dissipating the energy of the qubit system into the environment.
Mathematically, they are respectively described by vanishing of the
off-diagonal and diagonal elements of the reduced density matrix of $S$ .
The dissipation effect of imperfect isolation happens at the relaxation
time-scale $\tau _{rel}$. It is relatively easy to make systems having a
very large $\tau _{rel}$ and thus allowing a reasonable number of operations
to complete [10]. In contrast, the effect of decoherence, is much more
insidious [10, 37-42] because the coherence information leaks out into the
environment in a time scale $\tau _d$ much shorter than $\tau _{rel} $ as a
quantum system evolves [10, 37-42]. In fact, it is rather difficult to
realize a quantum system with the time scale $\tau _d$ for decoherence
smaller than that for the dissipation. For example, let us consider an
oscillator in a superposition of two coherent states separated by a distance 
$l$ from each other. Suppose they interact linearly with a bath of
oscillators at temperature T. Then the decoherence time for this system is
linearly proportional to the relaxation time with a rather small ratio $(%
\frac{\lambda _{them}}l)^2$ at high temperatures where $\lambda _{them}$ is
the thermal de Broglie wavelength [10, 37-42]. This implies that the qubits
decohere much faster than they dissipate. Thus, the sensibility of quantum
computation mainly depends on $\tau _d$rather than $\tau _{rel}$. For this
reason , the present discussions in this paper only focus on the decoherence
problem.

The arrangement of this paper is as follows. In sec. 2, we consider the
influence of environment on Shor's prime factoring algorithm; From sec. 3 to
sec.7 we consider decoherence problem caused by environment with concrete
models; In sec. 9, we discuss the universality of environment in the weakly
coupling limit; Finally in sec. 8 we draw some conclusions.\qquad

\qquad

\begin{center}
{\bf 2. Decoherence in Shor Factoring Algorithm}
\end{center}

\quad

In this section we illustrate the possible influence of environment on the
validity of a quantum algorithm through an example, Shor's prime
factorization algorithm . In this case, we recall, the so called quantum
computer has two registers in use. According to Shor's method [6], to factor
a number $n$ one should first of all choose a number $x$ . Then the first
step is to put the first register in the uniform superposition of q states $%
|a\rangle (a=0,1,....q-1)$ and the second one in a single state $|0\rangle .$
This leaves the machine in the state 
$$
|\phi (0)\rangle =\frac 1{\sqrt{q}}\sum_{a=0}^{q-1}|a\rangle \otimes
|0\rangle \eqno (2.1) 
$$
Next, one computes $x^qmod.(n)$ in the second register, leaving the machine
in the state 
\[
|\phi (t)\rangle =\frac 1{\sqrt{q}}\sum_{a=0}^{q-1}|a\rangle \otimes
|x^amod.(n)\rangle 
\]
$$
\equiv \frac 1{\sqrt{q}}\sum_{a=0}^{q-1}|a,x^amod.(n)\rangle \ \eqno(2.2) 
$$
Then one performs a Fourier transform $A_q$ on the first register. This
leaves the machine in the state 
$$
|\phi _F(t)\rangle =\frac 1q\sum_{a=0}^{q-1}\sum_{c=0}^{q-1}\exp [\frac{2\pi
iac}q]|c,x^amod.(n)\rangle \eqno (2.3) 
$$
Finally one observes the machine. One easily finds that the probability that
the machine ends in a particular state $|c,x^kmod.(n)\rangle \equiv
|c,k(x;n)\rangle $ is 
$$
p(c,k)=\frac 1{q^2}|\sum_{a:x^a=x^kmod.(n)}^{q-1}\exp [\frac{2\pi iac}q]|^2%
\eqno (2.4) 
$$
Shor shows that if $c$ lies in a particular region one can determine a
nontrival factor of $n$ from the value of $c$. Denote the one-try-success
probability of this method by $p_s$. Then one has the following result: 
$$
p_s\geq r\phi (r)p(c,k)\geq 1/\ln n\geq 1/3\ln n\eqno (2.5) 
$$
where $r$ is the least integer such that $x^r\equiv 1(mod.n)$ and $\phi $ is
Euler's totient function.

Now let us take the influence of environment into account to some extent.
Asume that the environment is comprised by $N$ particles. In this case we
denote by $\phi ^{\prime }(0),\phi _F^{\prime }(t),p^{\prime }(c,k)$ and $%
p_s^{\prime }$ the correspondences of $\phi (0),\phi _F(t),p(c,k)$ and $p_s$
respectively. Then we have 
$$
|\phi ^{\prime }(0)\rangle =\frac 1{\sqrt{q}}\sum_{a=0}^{q-1}|a\rangle
\otimes |0\rangle \otimes |e\rangle \eqno (2.5) 
$$
where $|e\rangle =|e_1\rangle \otimes |e_2\rangle \otimes ...\otimes
|e_N\rangle $ is the initial state of the environment without correlation
with the state of the machine. Here, $|e_k\rangle (k=1,2,...,N)$ denotes the
initial states of individual particle comprising the environment .
Accordingly, 
\[
|\phi ^{\prime }(t)\rangle =\frac 1{\sqrt{q}}\sum_{a=0}^{q-1}|a\rangle
\otimes |x^amod.(n)\rangle \otimes |e[a]\rangle 
\]
$$
\equiv \frac 1{\sqrt{q}}\sum_{a=0}^{q-1}|a,x^amod.(n)\rangle \otimes
|e[a]\rangle \eqno (2.6) 
$$
where $|e[a]\rangle =U_a(t)|e\rangle $ and $U_a(t)$ is the effective
evolution operator of the environment correlated with the state $|a\rangle .$
For simplicity we do not consider the influence of environment in the
process of the Fourier transform $A_q$.Thus we have 
\[
|\phi _F^{\prime }(t)\rangle =\frac 1q\sum_{a=0}^{q-1}\sum_{c=0}^{q-1}\exp [%
\frac{2\pi iac}q]|c,x^amod.(n)\rangle \otimes |e[a]\rangle 
\]
As the only difference between the present model and the original one is the
involvment of the environment variables $|e[a]\rangle $ in the entanglement,
to proceed along with the discussion we should consider the reduced density
matrix 
\[
\rho (t)=Tr_e(|\phi _F(t)\rangle \langle \phi _F(t)|)= 
\]
\[
\ \frac 1{q^2}\sum_{a=0}^{q-1}\sum_{c=0}^{q-1}\sum_{a^{\prime
}=0}^{q-1}\sum_{c^{\prime }=0}^{q-1}\exp [\frac{2\pi i(ac-a^{\prime
}c^{\prime })}q]\times 
\]
$$
\langle e[a^{\prime }]|e[a]\rangle |c,x^amod.(n)\rangle \langle c^{\prime
},x^{a^{\prime }}mod.(n)|\eqno (2.8) 
$$
Here we have traced over the environment variables. We notice that the
contribution of environment is given by the transition matrix element

$$
F(a,a^{\prime })\equiv \langle e[a^{\prime }]|e[a]\rangle =\ \langle
e|U_{a^{\prime }}^{\dagger }(t)U_a(t)|e\rangle \eqno (2.9) 
$$
$F(a,a^{\prime })$ is usually called decohering factor. Now it directly
follows that

\begin{center}
\[
p^{\prime }(c,k)=Tr\{\rho (t)|c,k\rangle \langle c,k|\} 
\]
$$
=\frac 1{q^2}\sum_{a,a^{\prime }:x^a=x^kmod.(n)=x^{a\prime }}^{q-1}\exp [%
\frac{2\pi i(a-a^{\prime })c}q]F(a,a^{\prime })\eqno (2.10) 
$$
\end{center}

We are now in a position to consider two extreme cases. For the first case,
suppose that the qubit system is completely isolated. In this case we have $%
U_{a^{\prime }}=U_a$ for $a^{\prime }\neq a$, so $\langle e[a^{\prime
}]|e[a]\rangle =1$ . As a result we get 
\[
p^{\prime }(c,k)=\frac 1{q^2}|\sum_{a:x^a=x^kmod.(n)}^{q-1}\exp [\frac{2\pi
iac}q]|^2 
\]
$$
=p(c,k)\eqno (2.11) 
$$

For the second case, suppose that the environment causes a complete
decoherence. If we indexed the elements of the reduced density matrix by $a$
and $a^{\prime }$, then this means that its off-diagonal elements vanish.
Such a case has been formulated in the dynamic theory of quantum measurement
as a consequence of a certain factorizable structure of the effective
evolution operator of environment[30-36]. In fact, if $U_a$ can be
factorized as 
$$
U_a(t)=\prod_j^NU_a^j(t)\eqno (2.12) 
$$
where $U_a^j(t)$ only concerns the j'th particle in the environment, the
decohering factor can be expressed as $N$-multiple product 
\[
F(a,a^{\prime })=\prod_j^N\langle e_j|U_{a^{\prime }}^{j\dagger
}(t)U_a^j(t)|e_j\rangle 
\]
$$
\ \equiv \prod_j^N|F^j(a,a^{\prime })\eqno (2.13) 
$$
of the decohering factors $F^j(a,a^{\prime })=\langle e|U_{a^{\prime
}}^{j\dagger }(t)U_a^j(t)|e\rangle $ with norms less than unity. In the
macroscopic limit $N\rightarrow \infty ,$ it is possible that $F(a,a^{\prime
})$ $\rightarrow 0,$ for $a^{\prime }\neq a$, namely,$\langle e[a^{\prime
}]|e[a]\rangle =\delta _{aa^{\prime }}.$ Then we have 
\[
p^{\prime }(c,k)=\frac 1{q^2}[(q-1-k)/r] 
\]
$$
\ \leq \frac 1{q^2}\frac qr=\frac 1{qr}\eqno (2.14) 
$$
and 
\[
p^{\prime }(s)=r\phi (r)p^{\prime }(c,k)= 
\]
$$
\phi (r)/q\leq \phi (r)/n^2\leq 1/n\eqno (2.15) 
$$

In the remaining part of this section let us proceed on to discuss the
possible influence of environment on the efficiency of Shor's prime
factorization algorithm.Generally speaking, a deterministic algorithm is
said to be efficient if the number of the computation steps taken to excute
it increases no faster than a polynomial function of $\ln N$ where $N$ is
the input. For a randomized algorithm this definition should be modified to
fit in the probability character. Suppose the one-try-success probability of
an randomized algorithm $A$ is $s$, then $A$ is said to be efficient if $%
\forall \varepsilon >0,\exists \ p(x)$ such that $\forall N\cdot
(1-s)^{p(\ln N)}<\varepsilon ,$ where $p(x)$ is a polynomial. Obviously, the
polynomial $p(x)$ here should have real coefficients and satisfy $p(\ln N)>0$%
. All the polynomials appearing in the following are tacitly assumed to have
this property. It is also clear that in quantum computations all algorithms
should be randomized ones.

Let $A$ be a quantum algorithm. Suppose for an input $N$ the one-try-success
probability of $A$ is $f(N)$ where $f$ is a real continuous function defined
on the real line.Then we have the following lemma.\qquad

\qquad

{\bf Lemma}.{\it \ If there exists a polynomial }$p(x)${\it \ such that } 
$$
\lim_{N\rightarrow \infty }(1-f(N))^{p(\ln N)}<1\eqno (2.16)
$$
{\it then }$A${\it \ is efficient. Conversely, if for an arbitrary
polynomial }$p(x)${\it \ we have } 
$$
\lim_{N\rightarrow \infty }(1-f(N))^{p(\ln N)}\geq 1\eqno (2.17)
$$
$A${\it \ is not efficient.\qquad }

\qquad

{\bf Proof}. Let $p(x)$ be a polynomial such that $\lim_{N\rightarrow \infty
}(1-f(N))^{p(\ln N)}<1.$ Then $\forall \varepsilon >0,\exists \alpha
(\varepsilon )$ such that$~($ $\lim_{N\rightarrow \infty }(1-f(N))^{p(\ln
N)})^{\alpha (\varepsilon )}<\varepsilon .$ Namely, $\lim_{N\rightarrow
\infty }(1-f(N))^{p(\ln N)\alpha (\varepsilon )}<\varepsilon $. Defining 
\[
p^{\prime }(x)\equiv \alpha (\varepsilon )p(x)
\]
we come to the conclusion that there exists some $N_0$ such that $\forall
N>N_0$, $(1-f(N))^{p^{\prime }(\ln N)}<\varepsilon .$ It is now evident that
one can choose a suitable polynomial $q(x)$ such that $\forall N,$ $%
(1-f(N))^{q(\ln N)}<\varepsilon .$ This proves the first part of the lemma.

For the second part of the lemma, if the conclusion were not true, $\forall
\varepsilon >0,$there would exist a polynomial $p_\varepsilon (x)$ such that 
$\forall N,$ $(1-f(N))^{p_\varepsilon (\ln N)}<\varepsilon .$Thus we would
have $\lim_{N\rightarrow \infty }(1-f(N))^{p_\varepsilon (\ln N)}\leq
\varepsilon ,$leading to the contradiction $1\leq \varepsilon $. The lemma
is consequently proved.

Before concluding this section let us take $A$ to be Shor's prime
factorization algorithm and return to the above mentioned two extreme cases.
For the first case, we have $f(N)>1/3\ln N$. As a result, 
\[
\lim_{N\rightarrow \infty }(1-f(N))^{3\ln N}\leq 
\]
\[
\lim_{N\rightarrow \infty }(1-1/(3\ln N))^{3\ln N}
\]
$$
=1/e<1\eqno (2.17)
$$
So according to the lemma $A$ is efficient. For the second case,we have $%
f(N)\leq 1/N$. It is easy to prove $\lim_{N\rightarrow \infty }(1-1/N)^{\ln
^mN}=1$ for all ntegers $m$ so for all polynomials $p(x)$ $%
\lim_{N\rightarrow \infty }(1-1/N)^{p(\ln N)}=1.$Consequently, for all
polynomials $p(x)$ 
\[
\lim_{N\rightarrow \infty }(1-f(N))^{p(\ln N)}
\]
$$
\geq \lim_{N\rightarrow \infty }(1-1/N)^{p(\ln N)}=1\eqno (2.17)
$$
This means that the algorithm is no longer efficient in this case.

\qquad

\begin{center}
{\bf 3. Dynamic Model for Decoherence in Quantum Computation:}

{\bf Generalized Hepp-Coleman Approach}

\quad

{\bf \qquad }
\end{center}

In this section we begin to study decoherence problems caused dynamically by
environment. Since the influence of environment on quantum computations is
reflected in the decohering factor $\langle e[a^{\prime }]|e[a]\rangle $
caused by coupling, as is shown in the last section,our discussion will
concern the microscopic dynamics of the interaction between a qubit system
and the environment. It is modeled in terms of a generalization of the HC
model of the wave function collapse in quantum measurement.

In the generalized HC model [30-32 ], the environment E or the detector is
made up of $N$ particles and has the free Hamiltonian in a general form 
$$
\hat H_D=\sum_k^N\hat H_k\eqno (3.1) 
$$
where the single particle Hamiltonian $\hat H_k$ only depends on dynamical
variables $x_k$ (such as the canonical coordinate , momentum and spin ). Let 
$|n\rangle (n=1,2,...L)$ be states of the system (e.g., a quantum register)
corresponding to energy levels $E_n~(n=1,2,...M)$. The system S with
Hamiltonian 
$$
{\normalsize \hat H_s=\sum_{n=1}^LE_n|n\rangle {\langle }n|}\eqno (3.2) 
$$
interacts with the environment through the quantum non-demolition (QND)[43]
interaction 
$$
{\normalsize \hat H_I=}\sum_n\sum_jg_{n_{,j}}(x_j)|n\rangle {\langle }n|%
\eqno (3.3) 
$$
In a special realization for quantum computation, $|n\rangle $ may denote an
array of qubits, each of which has two states $|0\rangle $ and $|1\rangle ,$
through the definition $|n\rangle $ $=|n_{o,}n_{1,}n_{2,}...,n_{L-1}\rangle $
where the labels satisfy the unique binary representation 
$$
n=\sum_{i=0}^{L-1}n_i2^i(n_i=0,1).\eqno (3.4) 
$$
It should be emphasized that, for quantum measurement, the interaction
between S and E must be chosen to have the different strengths for the
different states of S, i.e., it is required that $g_{nj}\neq g_{mj}$ for $%
m\neq n$. In fact, the so-called measurement is a scheme to read out the
states of S from the counting number of detector such that different
counting numbers should correspond to different states of S. However, this
requirement of non-degeneracy is not necessary when we extend this
generalized HC model such that it is applicable in quantum computation.

If the coupling of the system to the environment is degenerate, namely, $%
g_{n,j}(x_j)=g_{m,j}(x_{j)}$ for certain $n\neq m,$ we can group the
coefficients of the interaction as follows 
\[
g_{1,j}=....=g_{d_{1,}j}\equiv \kappa _{1,j,} 
\]
\[
g_{d_1+1j}=....=g_{d_1+d_2j}\equiv \kappa _{2,j,} 
\]
\[
\cdot \cdot \cdot \cdot \cdot \cdot \cdot \cdot \cdot \cdot \cdot \cdot
\cdot \cdot \cdot \cdot \cdot \cdot \cdot 
\]

\[
g_{d_1+..+d_{q-1}+1j}=....=g_{d_1+...+d_qj}\equiv \kappa _{q,j} 
\]
$$
.......................................\eqno (3.5) 
$$
Correspondingly, the Hilbert space $V:\{|n\rangle |n=1,2,...L\}$of S is
decomposed into a direct sum 
\[
V_s=\sum_q\oplus V^q 
\]
of the subspaces

\[
V^1:\{|n=m\rangle \equiv |1,m\rangle |m=1,...d_1\} 
\]

\[
V^2:\{|n=m+d_1\rangle \equiv |2,m\rangle |m=1,...d_2\} 
\]

\[
................................... 
\]

\[
V^q:\{|n=d_1+..+d_{q-1}+m\rangle \equiv |q,m\rangle |m=1,...,d_q\} 
\]
$$
...................................\eqno (3.6) 
$$
Notice that $\kappa _{qj}\neq \kappa _{q^{\prime }j}$ for $q\neq q^{\prime
}. $ With the above decomposition of the Hilbert space, the interaction
Hamiltonian can be re- written as 
$$
H_I=\sum_{q,m}\sum_j\kappa _{q,j}(x_j)|q,m\rangle {\langle }q,m|\eqno (3.7) 
$$
We observe that the coupling has the same strength for the states belonging
to the same subspace $V^q$.

In some cases the above classification of state vectors is a reflection of
the structure of some irreducible representation of certain group chain $%
G\supset K$ where $G,K$ are such chosen that $H_I$ is $G-$invariant and $H_s$
is at most $K-$ invariant. For instance, consider the group chain $%
SO(3)\supset SO(2)$ . It defines the standard angular basis$|J,M\rangle $
through the Casimir operators $\hat J^2$ and $\hat J_3$ of $SO(3)$ and $%
SO(2) $. In this case, we can take $\hat H_s=H(\hat J^2,\hat J_3)$ to be the
Zeeman Hamiltonian in a central force field (not Coulomb field) if the
interaction $H_I$ =$H_I(\hat J^2)$.

A special cases of the above general discussion has already been given in
ref. [21]. They introduce a totally factorized interaction as of form $H_I=%
\hat Q\otimes \sum_{j=1}^Nf_j(x_j).$ Here $\hat Q$ is a system variable
commuting with the free Hamiltonian $\hat H_s$ of the qubit system and $x_j$
(j=1,2,....N) are the variables of the environment with the free Hamiltonian 
$\hat H_D=\sum_k^N\hat H_k.$ The Hilbert space $V_s$ for the system can be
spanned by $|q,m\rangle (m=1,2,...,d_q$ for a given $q$ ) , the common
eigenstates of $\hat Q$ and $\hat H_s$ labeled by $q$ and $m.$%
$$
\hat Q|q,m\rangle =e_q|q,m\rangle ,\hat H_s|q,m\rangle =E_{qm}|q,m\rangle %
\eqno (3.8) 
$$
In this case we have the direct sum decomposition $V_s=\sum_q\oplus V^q$
with the eigen-spaces $V^q=Span\{|q,m\rangle |m=1,2,...,d_q\}$.This special
interaction can be extended to a most general form with many system
variables $\hat Q_j(j=1,2..K)$ that cancel certain subspaces of the qubit
system simultaneously [21 ]. In the view point of group representation
theory, this generalization enjoys an elegant mathematical structure [22,
23].

\qquad

\begin{center}
{\bf 4. State Reduction in Time Evolution}
\end{center}

\quad

In this section it is shown that the above general structure of subspace
decomposition (3.5) indeed dynamically leads to a scheme grouping the states
of qubit system to avoid decoherence. Let $V_d=V_1\otimes V_2\otimes
....\otimes V_{N-1}\otimes V_N$ denote the direct product Hilbert space for
the environment. Here $V_k(k=1,2,...,N)$ denotes the Hilbert space of the
k'th particle comprising the environment. We will prove that, with the QN
interaction (3.6) any coherent superposition $\sum_mC_m|q,m\rangle $ of the
states belonging to the same subspace $V^q$ is decoherence free while the
coherent superposition $\sum_qD_q|q,m_q\rangle $of the states belonging to
different subspaces may experience a WFC or decoherence. For a given initial
state $|\sigma (0)\rangle =|\sigma _j(0)\rangle \in V_d,$ of the
environment, and a given initial state $|f(0)\rangle
=\sum_{m,q}C_m^q|q,m\rangle ,$ of the system , the general initial state of
the total system $|\Phi (0)\rangle =|f(0)\rangle $ $\otimes |\sigma
(0)\rangle $ will evolve into an entangling state 
\[
|\Phi (t)\rangle =\sum_q\sum_m^{d_q}\exp [-iE_{qm}t]C_m^q|q,m\rangle \} 
\]
$$
\otimes \prod_j\exp \{-it[\hat H_k+\kappa _{q,j}(x_j)]\}|\sigma _j(0)\rangle %
\eqno (4.1) 
$$
where $E_{qm}=E_{d_1+..+d_{q-1}+m},m=1,2,..d_q.$ Notice that the states
belonging to a subspace $V^q$ entangle with the environment through the same
factorized components 
\[
\prod_jU_j^q(t)|\sigma _j(0)\rangle \equiv 
\]
$$
\prod_j\exp \{-i[\hat H_k+\kappa _{q,j}(x_j)]t\}|\sigma _j(0)\rangle \eqno %
(4.2) 
$$
This kind of factorization structure in evolution of wave function is
crucial to decoherence or WFC [30-36]. In fact, the reduced density matrix
of the system at time t is 
\[
\rho (t)=Tr_d(|\Phi (t)\rangle \langle \Phi (t)|)= 
\]
\[
\ \sum_q\{\sum_m|C_m^q|^2|q,m\rangle \langle q,m|+ 
\]
\[
\sum_{m\neq .m^{\prime }}\exp [iE_{qm^{\prime }}t-iE_{qm}t]C_m^qC_{m^{\prime
}}^{q*}|q,m\rangle \langle q,m^{\prime }|\} 
\]
\[
\ +\sum_{q\neq q^{\prime }}\sum_{m.m^{\prime }}\exp [iE_{q^{\prime
}m^{\prime }}t-iE_{qm}t]C_m^qC_{m^{\prime }}^{q^{\prime }*}|q,m\rangle 
\]
$$
\ \langle q^{\prime },m^{\prime }|\prod_{j=1}^N\langle \sigma
_j(0)|U_j^{q^{\prime }\dagger }(t)U_j^q(t)|\sigma _j(0)\rangle \eqno (4.3) 
$$
where $Tr_d$ means taking partial trace over the variables of the
environment. From this expression we see that each off-diagonal element of $%
\rho (t)$ , labeled by $q$ and $q^{\prime }$, is accompanied by a decohering
factor 
\[
F_{q,q^{^{\prime }}}(N,t)=\prod_{j=1}^N\langle \sigma _j(0)|U_j^{q^\prime
\dagger} (t) U_j ^q(t)|\sigma _j(0)\rangle 
\]
$$
\equiv \prod_{j=1}^NF_{q,q^{^{\prime }}}^j(t)\eqno (4.4) 
$$
in the form of factorized function .Obviously, if the initial state $%
|f(0)\rangle $ belongs to a single subspace $V^q,$ then the terms
accompanied by $F_{q,q^{^{\prime }}}(N,t)$ do not appear. Thus the system
will remain in the pure state $\exp [-iH_st]|f(0)\rangle \langle f(0)|\exp
[iH_st]$ throughout the evolution process. This fact is significant for
developing schemes to carry out error free quantum computations. The
expression also manifests the happening of decoherence for those states
belonging to different subspaces $\{V^q\}.$

Next we consider the dynamic process of decoherence when a superposition of
states mixes the vectors belonging to different subspaces. Naively, as $%
F_{q,q^{^{\prime }}}(N,t)$ is a multiplication of $N$ factors $%
F_{q,q^{^{\prime }}}^j(t)$ with norms not larger than the unity, it may
approach zero in the macroscopic limit with very large $N$. To deal with
this problem precisely, we define an real number not less than zero 
$$
\Delta _j^{q,q^{\prime }}(t)=-\ln |F_{q,q^{^{\prime }}}^j(t)|\eqno(4.5) 
$$
Then the norm of the accompanying factor $F_{q,q^{^{\prime }}}(N,t)$ is
expressed as 
$$
|F_{q,q^{^{\prime }}}(N,t)|=\exp [-\sum_{j=1}^N\Delta _j^{q,q^{\prime }}]%
\eqno(4.6) 
$$
Obviously, the series $\sum_{j=1}^\infty \Delta _j^{q,q^{\prime }}(t)$ $\geq
0$ since each term is not less than zero. There are two cases in which the
accompanying factor $F_{q,q^{^{\prime }}}(N,t)$ approaches zero in the
macroscopic limit with very large $N$. The first case is that the series $%
\sum_{j=1}^\infty \Delta _j^{q,q^{^{\prime }}}(t)$ diverges on $(0,\infty ]$%
. The second case is that the series converges to a monotonic function which
approaches zero as $N\rightarrow \infty .$ Therefore, it is possible that 
\[
\rho (t){}{}{\rightarrow }\sum_q\{\sum_m|C_m^q|^2|q,m\rangle \langle q,m|+ 
\]
$$
\sum_{m\neq .m^{\prime }}\exp [iE_{qm^{\prime }}t-iE_{qm}t]C_m^qC_{m^{\prime
}}^{q*}|q,m\rangle \langle q,m^{\prime }|\}\eqno(4.7) 
$$
as $N\rightarrow \infty .$ In the next section, some examples will be
presented to illustrate the above mentioned circumstances explicitly.

\qquad

\begin{center}
{\bf 5.Dynamic Decoherence in Environment Consisting of }${\bf N}${\bf \
Two- Level Subsystems}
\end{center}

\quad

To make a deeper elucidation of the above mentioned quantum dynamic
mechanism of decoherence in quantum computation and the relevant scheme of
grouping the states of a qubit system to be prevented from decoherence, in
this section, we model the environment as consisting of $N$ two level
subsystems. We recall that Caldeira and Leggett [37] have pointed out that
any environment weakly coupling to system may be approximated as a bath of
oscillators . On the condition that ``any one environmental degree of
freedom is only weakly perturbed by its interaction with the system'', they
have also justified describing the influence of environment by a coupling
linear in the bath variables up to the first order perturbation. With this
justification, we observed that any linear coupling only involves the
transitions between the lowest two levels (ground state and the first
excitation state) of each harmonic oscillator in environment though it has
many energy levels. Therefore in this case we can also describe the
environment as a combination of many two level subsystems without loosing
generality. In fact, for quantum computation, Unruh[11] and Palma et al [12]
have considered the harmonic oscillator environment. Their model is
equivalent to one introduced for the WFC in quantum measurement by Sun et.al
[32, 31,3 4] . A similar model has also been touched by Leggett et al
[37,38]and Gardiner[39] in studying the tunneling effect in a quantum
dissipative process. Here we choose equivalently the two level subsystem
model to manifest some characters independent of environment in the weakly
coupling limit and to demonstrate explicitly the qualitative calculation of
decoherence time through a sample example without quantum dissipation.

Let $|g_j\rangle $ and $|e_j\rangle $ be the ground and excited states of $j$
'th subsystem .We define the quasi-spin operators 
\[
\sigma _1(j)=|e_j\rangle \langle g_j|+|g_j\rangle \langle e_j| 
\]
\[
\sigma _2(j)=-i[|e_j\rangle \langle g_j|-|g_j\rangle \langle e_j|] 
\]
$$
\sigma _3(j)=|e_j\rangle \langle e_j|-|g_j\rangle \langle g_j|\eqno (5.1) 
$$
Then we introduce the Hamiltonian of the environment 
$$
H_e=\sum_{j=1}^N\hbar \omega _j\sigma _3(j)\eqno (5.2) 
$$
and the interaction coupling to a qubit system 
$$
H_I=f(S)\sum_{j=1}^N\hbar g_j\sigma _2(j)\eqno (5.3) 
$$
where $f(S)$ is function of the variable S of the qubit system.

In this section let us mainly focus on the simplest case where the system
consists of two qubits with the Hamiltonian

$$
H_s=\hbar \eta _1S_3(1)+\hbar \eta _2S_3(2)\eqno (5.4) 
$$
where $S(1)=\sigma _s\otimes 1,S_s(2)=1\otimes \sigma _s,(s=1,2,3)$ denote
spin operators acting on the first and the second qubits respectively; $%
\sigma _s(s=1,2,3)$ denoting the usual Pauli matrix. We consider the special
interaction given by

$$
f(S)=S_3(1)+S_3(2)\eqno (5.5) 
$$
It means that in our model the interaction has the same strength for
different states. This model is very simple, or even too simple in some
sense. But we would like to point out that the so called Free Hamiltonian
Elimination model in ref.[21] is substantially only a plain generalization
of the present example to the multi-pair case if one takes into account the $%
SU(2)$ rotation transformation.

Let $|1\rangle $ and $|0\rangle $ be the qubit states that satisfy $%
S_3|k\rangle =(-)^{k+1}|k\rangle ,(k=1,0).$ With the chosen interaction
form, the Hilbert space, spanned by 
\begin{eqnarray*}
\{|1,1\rangle &=&|1\rangle \otimes |1\rangle ,|1,0\rangle =|1\rangle \otimes
|0\rangle , \\
|0,1\rangle &=&|0\rangle \otimes |1\rangle ,|0,0\rangle =|0\rangle \otimes
|0\rangle \}
\end{eqnarray*}
contains a null subspace $V^0$ of $H_I$ spanned by $|1,0\rangle $ and $%
|0,1\rangle $ $.$ Any superposition $|\phi (0)\rangle =A|1,0\rangle
+B|0,1\rangle $ in this subspace will preserve its purity in evolution
process though the system has interaction with the environment. Precisely,
the pure state $|\phi (0)\rangle $ $\langle \phi (0)|$ will evolve into the
pure state $U_0(t)|\phi (0)\rangle $ $\langle \phi (0)|U_0^{\dagger }(t)$
where $U_0(t)=\exp [-i\eta _1tS_3(1)-i\eta _2tS_3(2)]$ is the free evolution
operator of the qubit system. Physically, this fact implies that no useful
information leaks out of the system in the process and the coherence is
preserved. This analysis can be easily generalized to the many bit case
where the free qubit Hamiltonian takes the form $H_s=\sum_{k=0}^{L-1}\hbar
\eta _kS_3(k)$ and its interaction with the environment is determined by 
$$
f(S)=\sum_{k=0}^{L-1}\lambda _kS_3(k)]\eqno (5.6) 
$$
where $L$ is the number of qubits used and $S_s(k)=\stackrel{k-1times}{%
\overbrace{1\otimes \cdot \cdot \otimes 1}}\otimes \sigma _s\otimes 1\otimes
\cdot \cdot \cdot \otimes 1.$The different $\lambda _k^{\prime }s$ indicate
that each single qubit has a different coupling to the same environment. In
the Hilbert space of this L-qubit system with the basis 
\[
|q\rangle =|q_0\rangle \otimes ||q_1\rangle \otimes |q_2\rangle \otimes
...\otimes |q_{L-1}\rangle , 
\]
$$
q_k=0,1;k=0,1,2,...,L\eqno (5.7) 
$$
the subspace V$^\xi $ preserving coherence can be spanned by those basis
vectors $|q\rangle $ satisfying 
$$
\sum_{k=0}^{L-1}\lambda _k(-)^{q_k+1}=const.\xi \eqno (5.8) 
$$

Let us return to the two qubit example. If a superposition contains a vector
outside the decoherence free subspace, decoherence will happen in an
entanglement of system state with environment state. For example, if the
initial state $|\varphi (0)\rangle =C|0,0\rangle +D|1,1\rangle $ of the
system involves states not belonging to $V^0$ while the environment is
initially in the vacuum state $|0\rangle _e=|g_1\rangle \otimes |g_2\rangle
\otimes \cdot \cdot \cdot \otimes |g_N\rangle $ where $|g_j\rangle $is the
ground state of $j^{\prime }$th two level subsystem , the corresponding pure
state density matrix $|\varphi (0)\rangle $ $\langle \varphi (0)|\otimes
|0\rangle _e\otimes _e\langle 0|$ of the total system formed by the qubits
plus the environment will experience a unitary evolution to reach a pure
state $\rho _T(t).$ Its reduced density matrix 
\[
\rho (t)=Tr_e\rho _T(t)=|C|^2|0,0\rangle \langle 0,0|+|D|^2|1,1\rangle
\langle 1,1| 
\]
$$
+\lbrack CD^{*}\exp [2i(\eta _1+\eta _2)t]F(N,t)|0,0\rangle \langle 1,1|+H.c]%
\eqno (5.9) 
$$
is no longer pure because the environment state becomes correlated with the
system state. Here the decohering factor 
$$
F(N,t)\equiv \prod_{j=1}^NF_j(t)=\prod\limits_{j=1}^N\langle
g_j|U_{j1}^{\dagger }(t)U_{j0}(t)|g_j\rangle \eqno (5.10) 
$$
is determined by the effective evolution operators

\[
U_{j\alpha }(t)=\exp [-i\omega _j\sigma _3(j)t-i\xi _\alpha g_j\sigma
_2(j)t]\quad 
\]
$$
\xi _1=2,\xi _0=-2,(\alpha =0,1)\eqno (5.11) 
$$
corresponding to the qubit states $|0,0\rangle $ and $|1,1\rangle $
respectively. Using the formula $\exp [i\overrightarrow{\sigma }\cdot 
\overrightarrow{A}]=\cos A+i\overrightarrow{\sigma }\cdot \overrightarrow{n_A%
}\sin A$ for a given vector $\overrightarrow{A}$ of norm $A$ along the
direction $\overrightarrow{n_A},$ we get the explicit form of $U_{j\alpha
}(t)$ 
$$
U_{j\alpha }=\cos (\Omega _{j\alpha }t)-i\left[ \sigma _2(j)\sin \theta
_{j\alpha }+\sigma _3(j)\cos \theta _{j\alpha }\right] \sin (\Omega
_{j\alpha }t)\eqno (5.12) 
$$
where $\tan \theta _{j\alpha }=\frac{\xi _\alpha g_j}{\omega _j},\quad
\Omega _{j\alpha }=\sqrt{(g_j\xi _\alpha )^2+\omega _j^2}.$ Then, we get the
decohering factor $F(N,t)=\prod\limits_{j=1}^NF(j,t)$, which is an $N-$%
multiple product of the factors $F(j,t)=1-2\sin ^2\theta _j\sin ^2\Omega _jt$
of norm less than 1. Here, we have used the definitions $\tan \theta _j=%
\frac{2g_j}{\omega _j},\quad \Omega _j=\sqrt{4{g_j}^2+\omega _j^2}$for the
special labels $\xi _1=2,\xi _0=-2.$ Therefore, the temporal behavior of
decoherence is described by 
$$
|F(N,t)|=\exp \sum\limits_{j=1}^N\ln |1-8\frac{g_j^2}{\Omega _j^2}\sin
^2\left( \Omega _jt\right) |\eqno (5.13) 
$$
In the weakly coupling limit that $g_j\ll \omega _j$, we get

$$
|F(N,t)|=e^{-S(t)}\equiv \exp \left( -\sum\limits_{j=1}^N\frac{8g_j^2}{%
\omega _j^2}\sin ^2\left( \omega _jt\right) \right) \eqno (5.14) 
$$

A special case is that the subsystems constituting the environment are
identical and the environment has a constant discrete spectrum, i.e., $%
\omega _k=$constant $\omega $, $g_k=$constant $g$. In this case, the
off-diagonal elements with the factor $\exp \{-\frac{8Ng^2}{\omega ^2}\sin
^2\omega t\}$approach zero as $N\rightarrow \infty $ for all t except those
satisfying $\omega t=2k\pi (k=0,1,2...)$. For general information, one needs
a detailed analysis about the behavior of the series $S(t)$ for various
spectrum distributions of the environment. Of special interest is the case
with continuous spectrum. In such case $S(t)$ can be re- expressed in terms
of a spectrum distribution $\rho (\omega _k)$ as 
$$
S(t)=\int_0^\infty \frac 8{\omega _k^2}\rho (\omega _k)g_k^2\sin ^2\omega
_kd\omega _k\eqno (5.15) 
$$
Notice that, in the case of discrete spectrum, the distribution means a
degeneracy : there are $\rho (\omega _k)$ subsystems possessing the same
frequency $\omega _k$. From some concrete spectrum distributions,interesting
circumstances may arise. For instance, when $\rho (\omega _k)=\frac 1\pi
\gamma /g_k^2$ the integral converges to a negative number proportional to
time t , namely, $S(t)=-\gamma t$. This shows that the norm of the
decoherence factor is exponentially decaying and as $t\rightarrow \infty ,$
the off-diagonal elements of the density matrix vanish simultaneously!
Another example of continuous spectrum is the Ohmic type [37,38] $\rho
(\omega _k)=\frac{2\eta \omega _k^2}{\pi g_k^2},$ which leads to a diverging
integral $S(t)\rightarrow \infty $ for $t\neq 0.$ In conclusion, in the
present example, we can choose a suitable spectrum distribution of the
oscillators in the detector, such that the series $S(t)$ diverges to
infinity, or in other words, the dynamical evolution of the system plus
environment results in the complete decoherence in the reduced density
matrix of S independent of the temperature. However, it is only an accident
situation owing to the special choice of the initial state. For a general
initial state, we will see, the decoherence process indeed shows a
temperature independence.

\quad

\begin{center}
{\bf 6. Decoherence Time for L-Qubit System}

\quad
\end{center}

Usually, when the coherence of a quantum system develops a characteristic
decay proportional to a factor of the form exp$(-t/t_d),$ $t_d$, which
characterizes the speed of the decoherence or the transition of the system
from the quantum regime to the classical one, is called the decoherence
time. Its value depends on the physical feature of the quantum system and
their interaction with the environment. For a single qubit system some
numerical estimates of $t_d$ have been made by DiVincenzo[44] for several
physical realizations. It ranges from 10$^4$ s (for nuclear spins )to 10$%
^{-12}$ s (for the electron-hole excitation in bulk of a semiconductor). In
practice, to carry out a quantum computation, one needs a large number of
qubits, e.g., in Shor's algorithm factoring large number $n,$ $L\propto \ln
n.$ Accordingly, in the following we extend the dynamic analysis to show how
the speed of decoherence becomes larger as the number of qubits increases.

Let us consider the L-bit system coupling to the environment mentioned in
the last section. The interaction constants $\lambda _k$'s are chosen so
that eigen-values 
$$
\xi (q)=\sum_{k=0}^{L-1}\lambda _k(-1)^{q_k+1}\eqno (6.1) 
$$
are not degenerate for \{$q_k=0,1$\}.Starting from an initial state $%
|\varphi (0)\rangle =C|q\rangle +D|q^{\prime }\rangle ,$ where 
$$
|p\rangle =\prod_{k=1}^{L-1}\otimes |p_k\rangle ,p=q,q^{\prime }\eqno (6.2), 
$$
the initial pure state density matrix of the total system formed by the
qubits plus the environment will experience a unitary evolution to reach a
pure state $\rho _L(t).$

Imitating the calculation process in the last section, we can obtain the
reduced density matrix $\rho (t)=Tr_e\rho _L(t)$. Its off-diagonal elements
are proportional to the decohering factor 
\[
F_L(N,t)=\prod\limits_{j=1}^NF_L(j,t) 
\]
$$
\ \equiv \prod\limits_{j=1}^N\langle g_j|U_{jq}^{\dagger }(L,t)U_{jq^{\prime
}}(L,t)|g_j\rangle \eqno (6.3) 
$$
where 
$$
U_{jq}(L,t)\equiv \prod\limits_{j=1}^N\exp [-i\omega _j\sigma _3(j)t-i\xi
(q)g_j\sigma _2(j)t]\eqno (6.4) 
$$
Using the notions 
$$
\tan \theta _j(q)=\frac{\xi (q)g_j}{\omega _j}, \quad \Omega _j(q)=\sqrt{%
[g_j\xi (q)]^2+\omega _j^2}\eqno (6.5) 
$$
and the matrix representation of $U_{jq}(L,t)$, after straight calculation
we get 
\[
F_L(j,t)=\sin \theta _j(q)\sin [\Omega _j(q)t]\sin \theta _j(q^{\prime
})\sin [\Omega _j(q^{\prime })t] 
\]

\[
+\{\cos [\Omega _j(q)t]-i\cos \theta _j(q)\sin [\Omega _j(q)t]\} 
\]
$$
\times \{\cos [\Omega _j(q^{\prime })t]+i\cos \theta _j(q^{\prime })\sin
[\Omega _j(q^{\prime })t]\}\eqno (6.6) 
$$

Trivially, $F_L(N,t)$ becomes unity when $q=q^{\prime }.$ However, when $%
q\neq q^{\prime }$, in the weakly-coupling limit $g_j\ll \omega _j$, we have 
\[
\sin \theta _j(q)\simeq \theta _j(q) 
\]
\[
\cos \theta _j(q)\simeq 1-\frac 12\theta _j^2(q) 
\]
$$
\Omega _j(q)\simeq \omega _j\eqno (6.7) 
$$
Thus $F_L(j,t)\simeq 1-\frac 12\{\theta _j(q)-\theta _j(q^{\prime })\}^2\sin
^2(\omega _jt)+\frac i4\{\theta _j^2(q)-\theta _j^2(q^{\prime })\}\sin
(2\omega _jt).$ Since in such weak-coupling limit $\theta _j(q)\simeq \sin
\theta _j(q)\simeq \xi (q)\frac{g_j}{\omega _j},$ we obtain the decohering
factors 
\[
F_L(j,t)\simeq 1-\frac{g_j^2}{2\omega _j^2}\{\xi (q)-\xi _j(q^{\prime
})\}^2\sin ^2(\omega _jt) 
\]
$$
\ +\frac{ig_j^2}{4\omega _j^2}\{\xi _j^2(q)-\xi _j^2(q^{\prime })\}\sin
(2\omega _jt)\eqno (6.8) 
$$
Consequently 
$$
|F_L(N,t)|=\exp \{-[\xi (q)-\xi (q^{\prime })]^2\sum_{j=1}^N\frac{g_j^2}{%
2\omega _j{}^2}\sin ^2(\omega _jt)\}\eqno (6.9) 
$$

In summary, the temporal behavior of the decoherence is described by $%
F(N,t), $ and actually determined by $|F(N,t)|$ , which is of the form $\exp
[-S_L(t)]$. Here $S_L(t)=[\xi (q)-\xi (q^{\prime })]^2\sum_{j=1}^N\frac{g_j^2%
}{2\omega _j{}^2}\sin ^2(\omega _jt)$. For identical qubits $\lambda _k=$1$,$%
the fastest decoherence happens between the two initial states $|q\rangle
=|q_0=1\rangle \otimes |q_1=1\rangle \otimes \cdot \cdot \cdot \otimes
|q_{L-1}=1\rangle $ and $|q^{\prime }\rangle =|q_0=0\rangle \otimes
|q_1=0\rangle \otimes \cdot \cdot \cdot \otimes |q_{L-1}=0\rangle $ . In
this case $|F_L(N,t)|=\exp [-L^2S(t)].$ Thus for the instance with $%
S(t)=\gamma t,$ which is discussed in the last section, we have $%
|F(N,t)|=\exp [-L^2\gamma t]$ where $\gamma ^{-1}$is the decoherence time
for a single qubit. This shows that the characterized time of the fastest
decoherence happening in the L-qubit system is $L^2$ times of that of a
single qubit. This conclusion first obtained by Palma et.al.[12] is given
here in the framework of quantum dynamic model of decoherence.

\qquad

\begin{center}
{\bf 7.Temperature Dependence of Decoherence}

\quad
\end{center}

The above discussion about decoherence in quantum computation only concerns
the situation of zero temperature. In this section we consider the influence
of environment at a finite temperature. Suppose the initial state of the
total system is described by a density matrix $\rho \left( 0\right) =\rho
_s(0)\otimes \rho _b(0)$ where $\rho _s(0)$ =$|\phi (0)\rangle \langle \phi
(0)|$ is the density matrix of the system, while $\rho _b(0)$ is that of the
bath

\[
\rho _b(0)=\frac{\exp (-\beta \hat H_b)}{Tr_b\exp (-\beta \hat H_b)}%
=\prod\limits_{j=1}^N\rho _{jb}(0) 
\]
$$
\ =\prod\limits_{j=1}^N\frac{e^{-\beta \omega \sigma _3(j)}}{2\cosh \left(
\beta \omega _j\right) }\eqno (7.1) 
$$
with $\beta =1/(K_BT).$ For the initial state $|\phi (0)\rangle
=A|0,0\rangle +B|1,1\rangle ,$ we obtain the same decohering factor $%
F_2(N,t)=$ $\prod_j^N[1-2\sin ^2\theta _j\sin ^2(\Omega _j t)]$ which is
independent of temperature. This is due to the special choice of the initial
state with a certain permutation symmetry between $|1,0\rangle $ and $%
|0,1\rangle $. For a general initial state $|\varphi (0)\rangle =C|q\rangle
+D|q^{\prime }\rangle ,$ we can calculate the factor $F_L(N,t)=\prod%
\limits_{j=1}^NF_L(j,t)$ as follows 
\[
\ F_L(j,t)\equiv Tr_b[U_{jq^{\prime }}(L,t)\rho _{jb}(0)U_{jq}^{\dagger
}(L,t)] 
\]
\[
=\sin \theta _j(q)\sin [\Omega _j(q)t]\sin \theta _j(q^{\prime })\sin
[\Omega _j(q^{\prime })t]+ 
\]
\[
\cos [\Omega _j(q)t]\cos [\Omega _j(q^{\prime })t]+ 
\]
\[
\cos \theta _j(q)\sin [\Omega _j(q)t]\cos \theta _j(q^{\prime })\sin (\Omega
_j(q^{\prime })t) 
\]
$$
-\frac i2\tanh (\beta \omega _j)\sin (2\Omega _j(q)t)\{\cos \theta
_j(q)-\cos \theta _j(q^{\prime })\}\eqno (7.2) 
$$
Notice that the effect of finite temperature only appears in the imaginary
part of the decohering factor. In the weakly coupling limit, it is not
difficult to observe that $\ F_L(j,t)\simeq 1-\frac 12\{\theta _j(q)-\theta
_j(q^{\prime })\}^2\sin ^2[\omega _jt]\ \ \ -\frac i4\tanh( \beta \omega
_j)\sin [2\omega _jt][\theta _j^2(q)-\theta _j^2(q^{\prime })]$ or 
\[
F_L(j,t)\simeq 1-\frac{g_j^2}{2\omega _j^2}\{\xi (q)-\xi _j(q^{\prime
})\}^2\sin ^2(\omega _jt) 
\]
$$
\ \ \ +\frac{ig_j^2}{4\omega _j^2}\{\xi _j^2(q)-\xi _j^2(q^{\prime })\}\tanh
(\beta \omega _j)\sin (2\omega _jt)\eqno (7.3) 
$$
It reflects the novel fact that, in an environment weakly interacting with
the qubit system, the decoherence time do not depends on temperature as a
result of the temperature-independent norm of $F_L(j,t).$ Thus in this case
thermal fluctuation plays a role in quantum computation only through
affecting the phases of the off-diagonal elements of the reduced density
matrix.

\quad

{\bf 8. Universality of Environment in Weak Coupling Limit\quad }

\qquad

An environment surrounding a qubit system for quantum computation maybe very
complicated. Intuitively, the dynamic process of decoherence in quantum
computation should depend on the details of interaction between the qubit
system and the environment. So generally it seems impossible to control
decoherence in a qubit system. Nevertheless, one may well expect that in
some limit situations there exists certain universality in the dynamics of
interaction so that the physical parameters (such as the decoherence time
and decoherence factor) dominating a quantum computation process would not
depend on the detail of environment . For the tunneling problem in quantum
dissipation process, this kind of universality has been considered by
Caldeira and Leggett [37, 38] by modeling the environment as a bath of
harmonic oscillators with a linear coupling to the system. In this section,
we illustrate that, in the weakly coupling limit, the above results obtained
from the two-level subsystem model of environment coincide with those from
the harmonic oscillator model concerned in various quantum irreversible
processes, such as wave function collapse [32,34,31] and quantum dissipation
[37-42].

Let $a_i^{+}$and $a_i$ be the creation and annihilation operators for the
i'th harmonic oscillator in the environment. The Hamiltonian of the
environment takes the form $H=\sum_{j=1}^N\hbar \omega _ja_j^{+}a$ and its
interaction with the qubit system can be modeled as a linear coupling: 
$$
H_I=f(s)\sum_{j=1}^N\hbar g_j(a_j^{+}+a_j)\eqno (8.1) 
$$
where $f(s)$ is a linear or non-linear function of the qubit system variable 
$s$. Let the initial state of the qubit system $|\varphi (0)\rangle
=A|\alpha \rangle +B|\beta \rangle $ be a coherent superposition of two
eigenstates of $s,s|\alpha \rangle =\alpha |\alpha \rangle ,s|\beta \rangle
=\beta |\beta \rangle $ and let the environment be initially in the vacuum
state $|0\rangle _e=|0_1\rangle \otimes |0_2\rangle \otimes \cdot \cdot
\cdot \otimes |0_N\rangle $ where $|0_j\rangle $is the ground state of the
j'th single harmonic oscillator. The corresponding decohering factor $%
F(N,t)=\prod_{j=1}^N\ _h\langle 0|U_j^{^\beta \dagger }(t)U_j^\alpha
(t)|0\rangle _h\equiv \prod_{j=1}^NF_j(t)$can be obtained by solving the
Schrodinger equations of $U_j^\gamma (t)$ ($\gamma =\alpha .\beta $)
governed by the Hamiltonian of forced harmonic oscillator 
$$
H_{j\gamma }=\hbar \omega _ja_j^{+}a_j+f(\gamma )g_j(a_j^{+}+a_j)\eqno (8.2) 
$$
In fact, by the so called Wei-Norman algebraic expansion technique one has
the following explicit result [12,32,35]. 
\[
F(N,t)=\exp \{-[f(\alpha )-f(\beta )]^2\sum_{j=1}^N\frac{2g_j^2}{\omega
_j{}^2}\sin ^2(\frac{\omega _jt}2)\} 
\]
$$
\times \exp \{-i[f(\alpha )^2-f(\beta )^2]\sum_{j=1}^N\frac{g_j^2}{\omega
_j{}}[t+\frac{\sin (\omega _jt)}{\omega _j}]\}\eqno (8.3) 
$$
The decoherence time is decided by the real part of $F(N,t),$ which is the
same as in eq.(6.8) from the two level model of environment in the weakly
coupling limit. This can easily be seen if only one replaces $\frac{\omega _j%
}2$ in the above equation by $\omega _j.$ Thus in the weakly coupling limit
the differences among different models of environment are only reflected in
the imaginary parts of the decohering factors This simply implies that in
this case the details of environment does not affect the speed at which a
quantum system approaches the classical kingdom. But they do affect the
success probability of a quantum computation.

\qquad

\begin{center}
{\bf 9. A Discussion\quad }

\quad
\end{center}

We have seen from our model that for a quantum register with L qubits the
relevant coherence develops a characteristic decay proportional to $\exp
[-L^2/t_dt]$ where $t_d$ is the typical decoherence time for a single bit.
Thus if a quantum algorithm calls for K elementary computation steps and
each step takes time $\tau $ on the average in order that the algorithm
could be feasible we should have the condition 
$$
L^2\tau K<t_d\eqno (9.1) 
$$
Generally speaking, this would pose a strong restriction on L and K. We need
to develop proper quantum error correction schemes to cope with this
difficulty caused by decoherence , which is unavoidable in the quantum
kingdom. Along this line there has been some progress. Nevertheless there is
another severe problem which may endanger the assumed great utility of
quantum computers. In section 2 we have shown that environment may affect
the efficiency of a quantum algorithm. Although our discussion is not
sophisticated enough it indeed gives us a frustrating information. This
problem deeply rooted in the quantum kingdom seems to have been ignored. We
think it is now time to face it seriously.

\qquad

{\it This work is supported in part by special project of the NSF of China
and the National Excellence Youth Foundation of China}

\qquad

\begin{center}
{\bf References}

\qquad
\end{center}

\begin{enumerate}
\item[1]  D. Deutsch, Proc. R. Soc. Lond. A {\bf 400}, 97 (1985).

\item[2]  S. LIoyd, Science {\bf 261}, 1569 (1993).

\item[3]  A. Barenco, Contemp. Phys. {\bf 37}, 375 (1996).

\item[4]  C. H. Bennett, Phys. Today , {\bf 47(10)}, 24 (1995).

\item[5]  A. Ekert and R. Jozsa, Rev. Mod. Phys. 68, 733 (199G)

\item[6]  P. Shor, i{\it n Proceedings of the 35th Annual Symposium on
Foundations of Computer Science 1994} (IEEE Computer Society Press, Los
Alamitos, C.A., 1994), p.l24-134.

\item[7]  J. I. Cirac and P. Zoller, Phys. Rev. Lett. {\bf 74}, 4091 (1995).

\item[8]  Q. A. Turchette, C. J. Hood, W. Lange, H. Mabuchi and H. J. Kimble
Phys. Rev.Lett. {\bf 75}, 4710 (1995);

\item[9]  C. Monroe, D. M. Meekhof, B. E. King, W. M. Itano and D. J.
Wineland, Phys. Rev. Lett. {\bf 75}, 4714 (1995).

\item[10]  I. L. Chuang, R. Laflamme, P. W. Shor and W. H. Zurek, Science 
{\bf 270}, 1633 (1995).

\item[11]  W. G. Unruh, Phys. Rev. A {\bf 51}, 992 (1995).

\item[12]  G. M. Palma, K. A. Suominen, and A. K. Ekert, Proc. R. Soc.
London, A {\bf 452}, 567 (1996).

\item[13]  W.H.Zurek, Phys.Today, {\bf 44(10)}, 36 (1991).

\item[14]  P. W. Shor, Phys. Rev. A {\bf 52}, R2493 (1995).

\item[15]  A. M. Steane, Phys. Rev. Lett. {\bf 77}, 793 (1996).

\item[16]  A. R. Calderbank and P. W. Shor, Phys. Rev. A {\bf 54}, 1098
(1996).

\item[17]  R. Laflamme, C. Miguel, J. P. Paz, and W. H. Zurek, Phys. Rev.
Lett. {\bf 77}, 198 (1996).

\item[18]  C. H. Bennett, D. P. DiVincenzo, J. A. Smolin, and W. K.
Wootters, Phys. Rev. A{\bf \ 54}, 3824 (1996).

\item[19]  P. W. Shor and R. Laflamme, Phys. Rev. Lett. {\bf 78}, 1600
(1997).

\item[20]  H. F. Chau, Phys. Rev. A55, R839 (1997).

\item[21]  L.M. Duan and G.c.Guo,Phys. Rev. Lett. {\bf 79}, 1953 (1997).

\item[22]  P. Zanardi and M. Rasetti, Phys. Rev. Lett. {\bf 79}, 3306(1997).

\item  P. Zanardi and M. Rasetti, Lett.Math.Phys., in press (1998)

\item[24]  J.von Neumann, {\it Mathemstische Gruandlage de Quantumechanik},
(Berlin, Julius, 1932).

\item[25]  K. Hepp, Hev.Phys.Acta, {\bf 45}, 237 (1972).

\item[26]  J.S. Bell, Hev.Phys.Acta, {\bf 48}, 93 (1975).

\item[27]  M. Cini, Nuovo Cimento,{\bf 73B}, 27(1983).

\item[28]  M. Namik and S.Pascazio, Phys.Rev., A{\bf 44}, 39(1991).

\item[29]  H. Nakazato and S. Pascazo, Phys.Rev.Lett., {\bf 70}, (1993).

\item[30]  C.P.Sun, Phys.Rev.A, {\bf 48}, 878 (1993).

\item[31]  C.P.Sun, Chin. J. Phys., {\bf 32}, 7(1994)

\item[32]  C.P. Sun, X.X. Yi, and X.J. Liu, Fortschr.Phys., {\bf 43}, 585
(1995)

\item[33]  X.J.Liu and C.P.Sun, Phys.Lett.A, {\bf 198}, 371(1995)

\item[34]  C.P.Sun, X.X.Yi, S.Y.Zhao, L.Zhang and C.Wang, Quantum
Semiclass.Opt. {\bf 9},119(1997)

\item[35]  C.P. Sun, {\it \ General Dynamical Model of Quantum Measurement
and Its Application to quantum Zeno Effect,} in Proceedings of the 4th
Drexel Symposium On Quantum noninterability, Philadelphia 1994,
(International Press, 1996).

\item[36]  C.P. Sun,{\it Generalized Hepp-Coleman Models for Quantum
Decoherence as a Quantum Dynamic Process}, Quantum Coherence and
Decoherence, ed.by K.Fujikawa and Y.A.Ono,, pp.331-334,(AMsterdam: Elsevier
Sciense Press,1996)

\item[37]  A. O. Caldeira and A. J. Leggett, {\it Ann. Phys}. (N.Y.), {\bf %
149, }374(1983).

\item[38]  A.J. Leggett, S. Chakravarty, A.T. Dosey, M.P.A .Fisher and W.
Zwerger, Rev.Mod.Phys, {\bf 59}, 1-87(1987).

\item[39]  C. Gardiner, {\it Quantum Noise}, (Berlin, Springer,1991).

\item[40]  L.H.Yu and C.P. Sun, Phys.Rev.A, {\bf 49, }592(19940).

\item[41]  C.P.Sun and L.H. Yu, Phys.Rev.A, {\bf 51},1845(1995).

\item[42]  C.P.Sun, Y.B.Gao, H.F. Dong and S.R.Zhao, Phys.Rev.E. {\bf 57,}
in press, (1998)

\item[43]  V. Braginsky and Khalili, {\it Quantum Measurement}, ( Cambridge
Univ. Press, London 1992).

\item[44]  D.P. DiVincenzo, Science, {\bf 270}, 255(1995)
\end{enumerate}

\end{document}